# *A Deep Learning Technique using Low Sampling rate for residential Non Intrusive Load Monitoring*
*Ronak Aghera, Sahil Chilana, Vishal Garg, Raghunath Reddy*
*International Institute of Information Technology, Hyderabad*

## ABSTRACT

Individual device loads and energy consumption feedback is one of the important approaches for pursuing users to save energy in residences. This can help in identifying faulty devices and wasted energy by devices when left On unused. The main challenge is to identity and estimate the energy consumption of individual devices without intrusive sensors on each device. Non-intrusive load monitoring (NILM) or energy disaggregation, is a blind source separation problem which requires a system to estimate the electricity usage of individual appliances from the aggregated household energy consumption. In this paper, we propose a novel deep neural network-based approach for performing load disaggregation on low frequency power data obtained from residential households. We combine a series of one-dimensional Convolutional Neural Networks and Long Short Term Memory (1D CNN-LSTM) to extract features that can identify active appliances and retrieve their power consumption given the aggregated household power value. We used CNNs to extract features from main readings in a given time frame and then used those features to classify if a given appliance is active at that time period or not. Following that, the extracted features are used to model a generation problem using LSTM. We train the LSTM to generate the disaggregated energy consumption of a particular appliance. Our neural network is capable of generating detailed feedback of demand-side, providing vital insights to the end-user about their electricity consumption. The algorithm was designed for low power offline devices such as ESP32. Empirical calculations show that our model outperforms the state-of-the-art on the Reference Energy Disaggregation Dataset (REDD).

## INTRODUCTION

Currently, residential energy consumption (i.e., direct energy consumption at home) accounts for more than 20% of the total energy consumption worldwide, of which two-thirds is from non-OECD countries (IEA, 2017). Residential energy consumption is rapidly growing, with an increase of 33.7% observed in the past two decades. With Economic development, urbanization, and the improvement of people's living standards in non-OECD countries, residential energy demand is anticipated to further increase (O'Neill et al., 2012).The International Electrotechnical Commission (IEC) has stated that the intelligent and economic use of electricity, as the primary energy source, will be the most important factor in solving energy problems (IEC 2010). Thus, efficient and sustainable utilization of energy has been an important area of research in recent times. Appliance Load Monitoring helps to reduce energy wastage by creating awareness among users. The detailed consumption patterns also enable utility companies in effective load management. Literature shows that energy feedback information provided by smart meters can enable consumers to reduce consumption between 5% to 15% (Darby 2006).

Non-Intrusive Load Monitoring (NILM) or Energy disaggregation is a process of estimating individual appliance energy consumption from the aggregate load data obtained from the mains meter. NILM is cost-effective as compared to device-level monitoring. It helps in producing appliance wise electricity consumption which helps users to reduce their consumption or to help operators to manage the grid. It also helps to identify faulty appliances and to survey appliance usage behaviour.

The major challenge for NILM is building generalizable NILM models that perform accurately in real-time disaggregation. Appliance disaggregation usually involves four main steps: data collection, feature extraction, learning, and appliance power retrieval. Numerous variations in features and algorithms have led to the development of various NILM techniques. Recently deep learning has found wide acceptance in the area of machine learning. It has been used in natural language processing, speech recognition, computer vision, and other real-time applications. Deep learning techniques automatically learn feature representations from the data. In this paper, we would like to propose a deep learning-based NILM approach. The main contributions of our work are as follows:

1. We used deep learning techniques, mainly 1D Convolution Neural Networks (CNNs) and Long Short Term Memory (LSTM) networks to solve the NILM problem.
2. We have applied data pre-processing techniques available in NILMTK and used the REDD dataset (Kolter and Johnson, 2011) for experimental analysis and validation of our proposed approach.

The paper is organized as follows. Section 2 presents a current state-of-the-art of NILM. Section 3 discusses our proposed methodology in detail, Section 4 shows the experimental setup and data pre-processing and Section 5 provides the experimental results for validating our proposed approach. Finally, we conclude with the conclusions section.

## Literature review and Background:

In this section, we provide a review of existing NILM approaches. The first work on NILM was conducted by Hart (Hart 1992) in the 1990s. The NILM method based on high frequency power reading requires sophisticated hardware. The supervised NILM algorithms apply classifiers such as SVM, KNN, decision tree, etc for appliance identification. They are more accurate but require initial training data. Recently there is a growing interest in the application of deep learning techniques for NILM problems. The authors in (De Paiva Penha and Castro 2018) have proposed a CNN based technique to identify appliances. The data from a public database (Kolter and Johnson, 2011) collected at a frequency of 1 Hz was used in their study. The authors in (Çavdar and Faryad 2019) describe a hybrid model for energy disaggregation through deep feature learning (DFL) on the residential energy disaggregation dataset (REDD). They compared three different disaggregation methods namely the convolutional neural network, 1D CNN-RNN, and long short-term memory (LSTM) and showed that the proposed 1D CNN-RNN model was performing better than others.

A sequence-to-point learning technique in (Zhang et al. 2018) takes the input feature like a window of the mains and outputs a single point of the target appliance. A convolutional neural network is used to train the model to learn the signatures of the target appliances. The proposed neural network approach was applied to real-world household energy data and was shown to be better than some of the existing NILM techniques.

A state-of-the-art energy disaggregation based on Long Short-Term Memory Recurrent Neural Network (LSTM-RNN) model is described in (Kim et al. 2017). A novel signature was developed to improve classification performance. They validated the performance of the proposed model on UK-DALE (Kim et al. 2017) and REDD datasets and showed that their model outperforms the advanced models.

The authors in (Kelly and Knottenbelt 2015), apply three deep neural network architectures for energy disaggregation, namely LSTM, denoising autoencoders; and a network that regresses the start time, end time and average power demand of each appliance activation. The proposed three neural nets achieve better accuracy and recall than either combinatorial optimization or factorial hidden Markov models.

The above research feeds raw aggregated data directly to a single neural network as input and gives disaggregated data of each appliance as output. In this article, we proposed to use the separate model for identification and power retrieval i.e. 1D-CNN for appliance state identification and LSTM for active appliance power retrieval. We analyse the model empirically, showing that the network using separate model for identification and power retrieval outperforms the models using single model for both. Our proposed load identification algorithm is improved by defining a better architecture and more generalizable model.

**Artificial Neural Network**

An artificial neural network (ANN) is a directed graph where the nodes are artificial neurons and the edges allow information from one neuron to pass to another neuron (or the same neuron in a future time step). Usually, neurons are arranged into layers such that each neuron of layer m is connected to every neuron of layer m+1. The connection between two neurons is called weights. ANN learn by modifying these weights. ANN has an input and an output layer and in addition to these, any layer in between are called hidden layers. The ANN is a feed-forward network i.e. information flows from the input layer, through all hidden layers and to the output layer. The learning (updating the weights) of such kind of network is done during the backpropagation (Kelly and Knottenbelt 2015).

**Convolution Neural Network**

Convolution neural networks (CNN) are commonly used for image processing. It consists of three layers namely, Convolution layer, pooling layer and Fully connected layer. The architecture of CNN is based on the alteration of convolution layers and pooling layers (Hijazi, Kumar and Rowen 2015). The convolution layer consists of neurons that are responsible for extracting different sub-region resources from the input image. After the convolution layer, the pooling layer comes which reduces the number of connections to the following layers. We are using a Max-pooling layer. Max pooling layer returns the maximum values obtained in the filters. In the end, the fully connected layer connects all the neurons of the interior layer to the output neurons, which in turn represent the classes to be classified (Hijazi, Kumar and Rowen 2015). In computer vision, they use 2D images as inputs so they use 2D CNN. CNN captures small features better than the ANN or multi-layer perceptron. To capture the small important features of each appliances we have used the CNN. In this paper, we are using 1D CNN because our input data is a time-sequential data (1D data).

**Recurrent Neural Network:**

From the previous discussion, we saw that the convolution neural network is a feed-forward network which maps from the input vector to a single output vector. In this neural network, when it is fed with a new input, the memory of previous input is removed. Recurrent neural networks (RNNs) allow cycles in the network graph such that the output from neuron i in layer l at time step t is fed via weighted connections to every neuron in layer l (including neuron i) at time step t + 1 (Kelly and Knottenbelt 2015). This allows RNNs to map from the entire history of the inputs to an output vector. This makes RNNs well suited to sequential data. RNNs can suffer from "vanishing gradient" problem (Hochreiter and Schmidhuber 1997) where gradient information disappears or explodes as it back propagates. The one solution for this problem is "Long short term memory" (LSTM) architecture (Hochreiter and Schmidhuber 1997). LSTM uses a "memory cell" with a gated input, gated output and gated feedback. The intuition behind LSTM is that it is a differentiable latch (where a latch is the fundamental unit of a digital computer's RAM). LSTMs have been used with success on a wide variety of sequence tasks including automatic speech recognition (Graves and Jaitly 2014; Hochreiter and Schmidhuber 1997) and machine translation (Sutskever, Vinyals and Le 2014).

## Proposed Methodology:

**Problem Statement:**

The NILM dataset is time sequential data. Thus, models which are dealing with sequential data are generally used for NILM. The deep learning models which are currently used are CNN, RNN, and LSTM (Zhang et al. 2018; Kelly and Knottenbelt 2015; Chen et al. 2018; Zhang and Yang 2019; De Paiva Penha and Castro 2017) or it's variant. However, these models mainly have four problems to classify or regress the appliances on low-frequency data. The problems are listed below :

1. Requires transient or high-frequency steady-state features which require additional hardware equipment installed (Chang and Yang 2009) which further leads to high hardware cost.
2. Large number of parameters for each model.
3. Lower accuracy on multistate appliances like washing machines and dishwashers.
4. Requires large storage space for data.

In this paper, we are considering the above four problems and proposed the deep learning technique for low sampling data. The approach is the combination of the 1D-CNN (convolution neural network) and LSTM (Long short term memory). 1D-CNN is used for identifying the operation state of the appliance and LSTM is used for power retrieval of that appliance according to its consumption pattern. To capture all possible dependencies, we have trained 1D-CNN and

LSTM models separately for all the devices. The proposed method is suitable for the low-frequency data and requires low computational power once the model is trained. Once the neural network is trained, it does not need ground truth data from each house to predict. We have developed a series of experiments on the REDD dataset and tested the performance of the proposed method. Our main goal is to deploy the trained model on low power embedded system with TensorFlow library such as ESP32 and predict the real-time electricity consumption without deep neural network web service.

**Proposed Approach:**

For training our approach includes three steps, the same as (Yuan et al. 2019). First, we extract the power consumption pattern signature of each appliance which will be used to estimate the power value. The aggregation power is used to train the 1D-CNN network to identify the operating state of each appliance and at last, we used the active operating states for training LSTM for retrieving the power value. Figure 1 shows the flow chart of the proposed method. The details of the method are given in the following section 4.

For testing we feed aggregated power window to the 1D-CNN trained model and retrieve appliance operating state of each appliance and at last, we feed the operating state from 1D-CNN to LSTM to get the power value of each appliance. Deployment of the experiment is describe in implementation section.

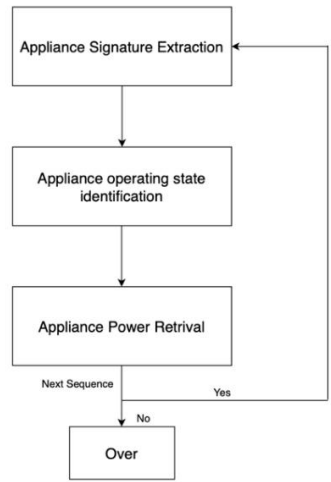

Figure 1: Flowchart of the proposed method

**Appliance Signature Extraction:**

Every appliance shows a unique power consumption pattern while operating. In this section, we will extract the power pattern signature of each appliance which will be used to predict power consumption. We extract the continuous sequence of power consumption of the appliances when their operation state is active. These extracted sequences are further used to train the LSTM model.

**Appliance operating state identification:**

In this paper, we are using the sequence to point (Zhang et al. 2018) approach for learning. It trains a neural network to only predict the midpoint of the window, not the entire window. The idea is that the input of the network is a mains window $X_{t:t+W-1}$, and the output is the midpoint element $y_\tau$ of the corresponding window of the target appliance, where $\tau = t + \lfloor W/2 \rfloor$. This type of method is called a sequence-to-point learning method which is widely applied for modelling the distributions of speech and image (Sainath et al. 2015). This method assumes that the mid-point element is represented as a non-linear regression of the mains window(Zhang et al. 2018). The intuition behind this assumption is that we expect the state of the midpoint element of that appliance to relate to the information of mains before and after that midpoint. The paper (Zhang et al. 2018) shows explicitly in their experiments that the change points (or edges) in the mains are the features that the network uses to infer the states of the appliance. We have tested the window length of 15, 20, 25, and 30 and find the F1- score of each. Window length of 20 gives the best result i.e. 0.91 F1-score on all the houses for all 4 appliances. Thus, we are using the sequence to point approach by providing the input of an array of length 20 to the network and predict the operating state corresponding to the 10th value of the array. In this paper, we are using the each 1D-CNN model for identifying an operating state of each appliance i.e. every appliances have separate 1D-CNN model. We are using categorical cross-entropy function as a loss function to get better results. Figure 2 shows the architecture of the 1D-CNN used in this paper.

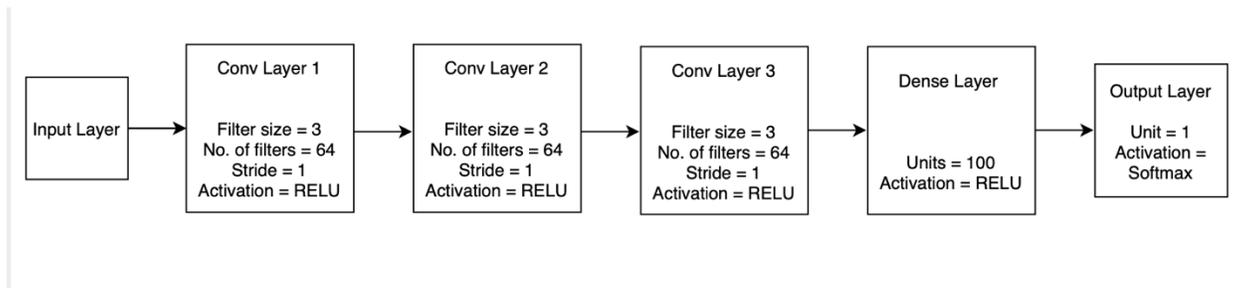

Figure 2: Architecture of 1D-CNN network in our model.

**Appliance power retrieval:**

We are using a sequential LSTM for predicting the power consumptions because the power data is time-series data. The input of the LSTM would be the past five states of the appliances and output will be the power value for the present time step. We are using the Rectified Linear unit (RELU) as an activation function at each layer and mean square error as a loss function in our model. We are training each model for each appliance. Below is the architecture of LSTM used in this paper.

LSTM model:

1. Input layer ( length =5)
2. Sequential LSTM layer (N= 50 , activation = RELU)
3. Sequential LSTM layer (N= 50 , activation = RELU)

4. Dense Layer
5. Loss Function: Mean Square Error (MSE)

**Combined schematic of whole model:**

The figure shows that the model takes the input of n ( 1*20) aggregated power readings and fed to the 1D-CNN. 1D-CNN gives the operating state at each timestep i.e. n (1*1). This operating state is indexed based on the active time i.e. if the device is inactive, index would be 0, if the device is active and previous 4 timestep is also active, index would be 5. After indexing the data-pre-processing for LSTM is done. We are giving past 5 indexes to LSTM to predict power value at the current timestamp. The list of dimension n(1*1) is changed to n(1*5) and fed to the LSTM which returns the active power. Thus, we fed aggregated power and got the power consumption by appliance. We are building appliance identification models for each appliance type so unseen appliances will not identified.

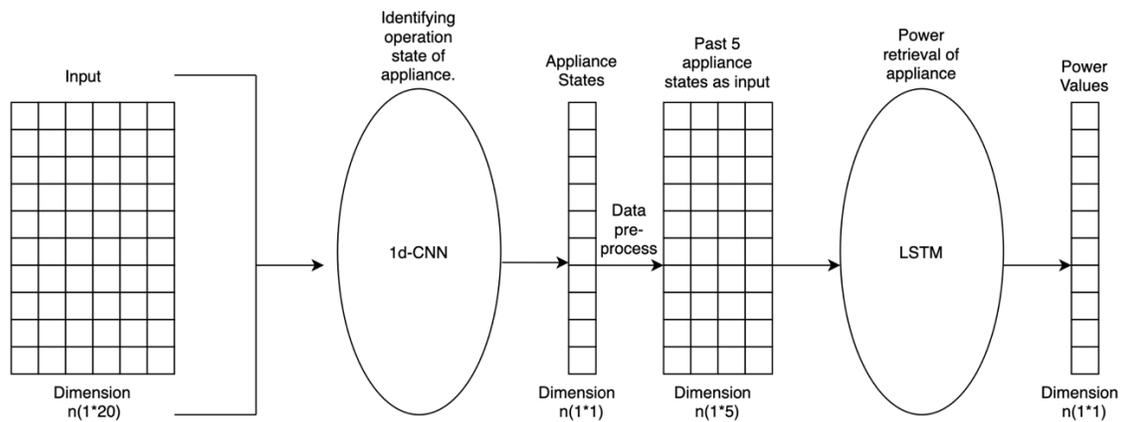

Figure 3: Pictorial representation of proposed method.

# Experimental setup and Data pre-processing:

**Dataset:**

In this study, we have used the reference energy disaggregation dataset (REDD) to verify the effectiveness of the proposed NILM. It is the most popular dataset for evaluating energy disaggregation algorithms as it contains both aggregate and sub-metered power data from six households. The dataset can be classified into three frequencies: low_freq, high_freq, and high_freq raw. High_freq and high_freq raw dataset contains voltage and current waveform. In this paper, we are going to use low_freq data as it contains power reading at 1 hz in each houses. The readings are at appliance and main level which are recorded every 3 seconds and 1 second respectively (Kolter and Johnson 2011). The dataset has been downsampled to 1 min for our experiment. The activation of the selected appliance per house is shown in table-1. The network presented in this paper is a combination of 1D-CNN and LSTM. We are training one network per appliance so that the noise of one appliance doesn't affect the other. Data preprocessing and experimentation is described below.

Table 1. Number of activations of appliances per house.

| House No. | Main meter | Refrigerator | Dishwasher | Microwave | Washing Machine |
|---|---|---|---|---|---|
| 1 | 26297 | 6561 | 1034 | 487 | 534 |
| 2 | 20034 | 8739 | 247 | 134 | 0 |
| 3 | 24601 | 8697 | 284 | 167 | 1036 |
| 4 | 27255 | 0 | 219 | 0 | 0 |
| 5 | 5216 | 1399 | 54 | 0 | 0 |
| 6 | 18550 | 8571 | 0 | 0 | 0 |

**Data Pre-processing:**

The Initial pre-processing of a dataset is done using the Non-intrusive load monitoring Toolkit (NILMTK) (Batra et al. 2014). The NILMTK is an open-source toolkit which is designed specifically to enable the comparison of energy disaggregation algorithms in a manner that is reproducible. The toolkit includes parsers for a range of existing data sets, a collection of pre-processing algorithms, a set of statistics for describing data sets, two reference benchmark disaggregation algorithms and a suite of accuracy metrics. (Batra et al. 2014).

In this paper, the dataset is down sampled from 3-sec data to 1 min data. All the gaps and NAN values in readings are detected and removed using NILMTK. After that, the longest continuous time sequence readings are identified with the help of a section function of NILMTK. Using the Electric.get_activations() method in NILMTK, appliance activations and their corresponding power consumptions are extracted. As we are dealing with the continuous time-series data so it is necessary to obtain the continuous-time sequence. Thus, we extract all the continuous-time sequences, convert the length of the sequences into multiples of 20 by adding zeros or removing some values and store it separately. The training and testing data will be taken from these time sequences. In this paper, we train 2 models for each appliance. Data pre-processing for each is explained below.

**Data pre-processing for 1D CNN:** The 1D-CNN model is used for identifying the operating state of the appliances (0 or 1). 1D-CNN uses the sequence to point approach as mentioned in the above section. It takes an input array of dimension n(1*20) where n is the number of timestamps and gives output n(1*1). For the endpoint of the full input sequence $X = (x_1, x_2, ..., x_t)$ and $y = (y_1, y_2, ..., y_t)$ we have added zeros at both the ends. The number of zeros to be added are calculated by [Window length /2] i.e. 10 zeros at beginning and 10 zeros at end in our case. After adding the zeros, we slide a window of 20 from the starting point of each continuous sequence till the end and store the array in X and their corresponding output in Y.

**Data pre-processing for LSTM:** The output from the 1D-CNN is the operating state of the appliances at the corresponding time step. After the pre-processing of the operating state we are feeding them to LSTM to predict the power value of that appliance at that time step. 1D-CNN gives the operating state at each timestep i.e. n (1*1). This operating state is indexed based on the active time i.e. if the device is inactive, the index would be 0, if the device is active and previous 4 timestep is also active, index would be 5. The power value and index is stored as shown below:
Power reading of refrigerator: [ 148, 135, 129, 127, 127, 125]
Power reading stored: [ (1,148), (2,135), (3,129), (4,127), (5,127), (6,125)]

To learn the power consumption pattern the LSTM takes the input array of past 5 appliance state indexes and gives power value as output. For training purpose, the length of input and output arrays is 5, the power at that time respectively is shown below:

Input (X): ——————————> Output (Y)
[0,0,0,0,0]————————————> 148
[0,0,0,0,1]————————————> 135
[0,0,0,1,2]————————————> 129
[0,0,1,2,3]————————————> 127
[0,1,2,3,4]————————————>127
[1,2,3,4,5]————————————>125

The window of length 5 is moved on operating index readings and fed to the LSTM to learn the power consumption pattern of each appliance. X= n(1*5) and y = n(1*1). Note that we are using only power readings when the appliance is active. For inactive power readings, we are taking the mean of the inactive power value of appliances from the dataset.

**Performance evaluation metrics:**

The performance evaluation of disaggregation algorithms has been one of the main challenges of NILM. It is important to produce an appropriate evaluation metrics. However, much of the literature focuses on the accuracy of on/off detection (Kelly and Knottenbelt 2015; Zhang and Yang 2019; De Paiva Penha and Castro 2017), and only a few studies considered the retrieving appliances' consumption information. We adopted a set of metrics introduced in the literature (Zhang et al. 2018; Chen et al. 2018) for power value and for appliance operating state identification we used metrics introduced in the literature (Zhang et al. 2018; Kelly and Knottenbelt 2015; Chen et al. 2018; Zhang and Yang 2019; De Paiva Penha and Castro 2017). All these metrics are explained in (Handelman et al. 2019) from machine learning point of view.

The metrics we are using are:

TP = number of true positive; TN = number of true negative; FP = number of false positive; FN = number of false negative;
Precision = TP / TP + FN; Recall = TP / TP + FP;
F1 score = 2*(Precision*recall) / (Precision + recall); Accuracy = TP + TN / P + N;
Mean Absolute Error = $(\sum_{t=1}^{t} |y_{pred} - y|) / t$; Mean Square Error = $(\sum_{t=1}^{t} |y_{pred} - y|)^2 / t$.

**Implementation:**

The programming language that has been used for writing the code is python. The Keras library is used for 1D-CNN and LSTM. For data pre-processing we have used Pandas, Numpy and NILMTK. The models have been trained on the 12 hr free GPUs named Tesla K80 which are powered by Google Colaboratory using the TensorFlow backend. Google Colab is a research tool based on Jupyter notebook environment for machine learning (ML) research and education. No setup is required to use a Jupyter notebook environment (Randles et al. 2017). The inference is much cheaper when these DNNs are trained; it takes approximately a processing second per network of DNN for a week of aggregate data on a GPU. The neural nets learn efficiently if the input data have mean between 0 to 1. So we normalized our data using MinMax normalisation. MinMax normalisation is done using the following formula:

$$x = x - xmin / xmax – xmin$$

We have tested our algorithm in two different ways:

1. Training on some houses and testing on the unseen house.
2. Training and testing on the same house.

**Training on some houses and testing on the unseen house:**

For testing the generalizability of the proposed algorithm we require an adequate number of appliance activation for training data. Thus, we are choosing the houses based on the adequate number of the training samples for training. Table 2 lists the houses that are used for training and testing. In the experiment, the appliances chosen to disaggregate are refrigerator, dishwasher, and microwave. These appliances are the most common household appliances in all six houses of REDD dataset, contribute the most towards the household's power consumption of REDD dataset. Mean square error (MSE) and mean absolute error (MAE) are used to evaluate the performance of disaggregation. The results of our proposed methods are compared with seq2seq (Kelly and Knottenbelt 2015), seq2point (Zhang et al. 2018), GLU-Res (Chen et al. 2018) and CNN algorithms (Zhang and Yang 2019).

Table 2. Selected houses for training/testing.

| Appliance | Training | Testing |
|---|---|---|
| Refrigerator | 2,3,5,6 | 1 |
| Microwave | 1,2 | 3 |
| Dishwasher | 1,2 | 4 |

**Training and testing on the same house:**

We have divided the data of each house into the ratio of 70:30 for training and testing. We have calculated precision, recall, F1 score and accuracy of proposed model in all the houses and compared the results with Deyvisan (De Paiva Penha and Castro 2017), seq2seq autoencoder (Kelly and Knottenbelt 2015), and seq2seq LSTM (Kelly and Knottenbelt 2015).

## Results and Conclusion:

### Experimentation and Results:

Below are the results of two experiments:

**Training on some houses and testing on unseen house:** We have compared the MAE and MSE on unseen houses with seq2seq (Kelly and Knottenbelt 2015), seq2point (Zhang et al. 2018), GLU-Res (Chen et al. 2018), and CNN (Zhang and Yang 2019). All these algorithms are trained and tested on the 1 or 6 sec frequency data and our algorithm is trained and tested on the 1 min (60 sec) frequency data. As shown in Table 3 our method outperforms the other 4 methods in power retrieval of microwave and dishwasher. The model reduces MAE and MSE by approximately 50% in microwave and dishwasher compared to the other four models. In the case of the refrigerator, the proposed model performed better than seq2seq (Kelly and Knottenbelt 2015) and seq2point (Zhang et al. 2018) but worse than GLU-Res (Chen et al. 2018) and CNN (Zhang and Yang 2019). Table 3 demonstrates how well the proposed approach performs on the unseen data. Thus, we can see that the proposed method has a capability for generalization.

Table 3. Disaggregation performance on houses not seen during training.

| Model | Frequency | Metrices | Refrigerator | Microwave | Dishwasher |
|---|---|---|---|---|---|
| Seq2seq | 6 sec | MAE | 30.6 | 33.3 | 19.5 |
|  |  | MSE | 2151.9 | 19292.8 | 14172.6 |
| Seq2point | 6 sec | MAE | 28.1 | 28.2 | 20.0 |
|  |  | MSE | 2393.9 | 17483.5 | 15891.3 |
| GLU-Res | 1 sec | MAE | 23.5 | 28.4 | 33.4 |
|  |  | MSE | 2197.4 | 25202 | 22301.1 |
| CNN | 1 sec | MAE | 21.8 | 18.3 | 22.3 |
|  |  | MSE | 1622.8 | 17037.9 | 18658.5 |
| Proposed_model | 1 min | MAE | 27.7 | 12.3 | 13.9 |
|  |  | MSE | 4486.7 | 9074.8 | 8144.5 |

The other models have not given the performance based on appliance state identification for unseen houses. Table 4 shows the identification performance on unseen house.

Table 4. Identification performance on unseen house.

| Appliance | Precision | Recall | F1 score | Accuracy |
|---|---|---|---|---|
| Refrigerator | 0.85 | 0.80 | 0.82 | 0.88 |
| Microwave | 0.61 | 0.80 | 0.69 | 0.99 |
| Dishwasher | 0.80 | 0.94 | 0.86 | 0.97 |

**Training and testing on the same house:** We have compared the operating state identification with Deyvisan (De Paiva Penha and Castro 2017), seq2seq autoencoder (Kelly and Knottenbelt 2015), and seq2seq LSTM (Kelly and Knottenbelt 2015). All these algorithms are trained and tested on the 1 or 6 sec frequency data and our algorithm is trained and tested on the 1 min (60 sec) frequency. As shown in Table 5, the proposed method outperforms the other three methods.

Table 5. Identification performance on houses seen during training.

| Paper | Precision | Recall | F1-score | Accuracy |
|---|---|---|---|---|
| Deyvisan | 0.82 | 0.84 | 0.82 | 0.82 |
| Seq2seq LSTM | 0.69 | 0.39 | 0.39 | 0.68 |
| Seq2seq Autoencoder | 0.80 | 0.58 | 0.55 | 0.91 |
| Proposed method | 0.91 | 0.91 | 0.91 | 0.97 |

The energy prediction on the same house is shown in Table 6:

Table 6. Disaggregation performance on houses seen during training.

| Appliance | MAE | MSE |
|---|---|---|
| Refrigerator | 21.0 | 1790.7 |
| Microwave | 12.1 | 5812.7 |
| Dishwasher | 9.8 | 3001.3 |
| Washing Machine | 18.8 | 17951.9 |

We are comparing the number of trainable parameters of our model with seq2seq (Kelly and Knottenbelt 2015), seq2point (Zhang et al. 2018), GLU-Res (Chen et al. 2018) and CNN (Zhang and Yang 2019). All these algorithms are trained on house 2 to 6 and tested on 1. As shown in Table 7, the number of trainable parameters of our model is smaller than other models.

Table 7. Size of the parameter of each models.

| Model | Number of trainable Parameter (in Millions) |
|---|---|
| Seq2seq | 29.8 |
| Seq2point | 29.2 |
| GLU-Res | 1.2 |
| CNN | 0.738 |
| Proposed method | 0.070 |

The results shown in the Table 3,5 and 7 show that our approach is able to correctly detect the operating state of appliances with a higher accuracy, have good generalizability and lower size of model than any other state-of-art. Thus, we can conclude that our model has a good capability for appliance identification and good generalizability.

F-1 score and accuracy for identification on seen houses is higher than the unseen houses as shown in Table 4 and 6, due to the lower MAE and MSE on seen houses than the unseen houses.

Figure 4 represents the disaggregation result of proposed model on 500 continues test points of house 1. It has 2 parts, first part is aggregated reading, and the second part is the comparison with predicted and actual disaggregated power values. We have considered refrigerator, microwave, dishwasher and washing machine for disaggregation. Each diagram represent the different appliances.

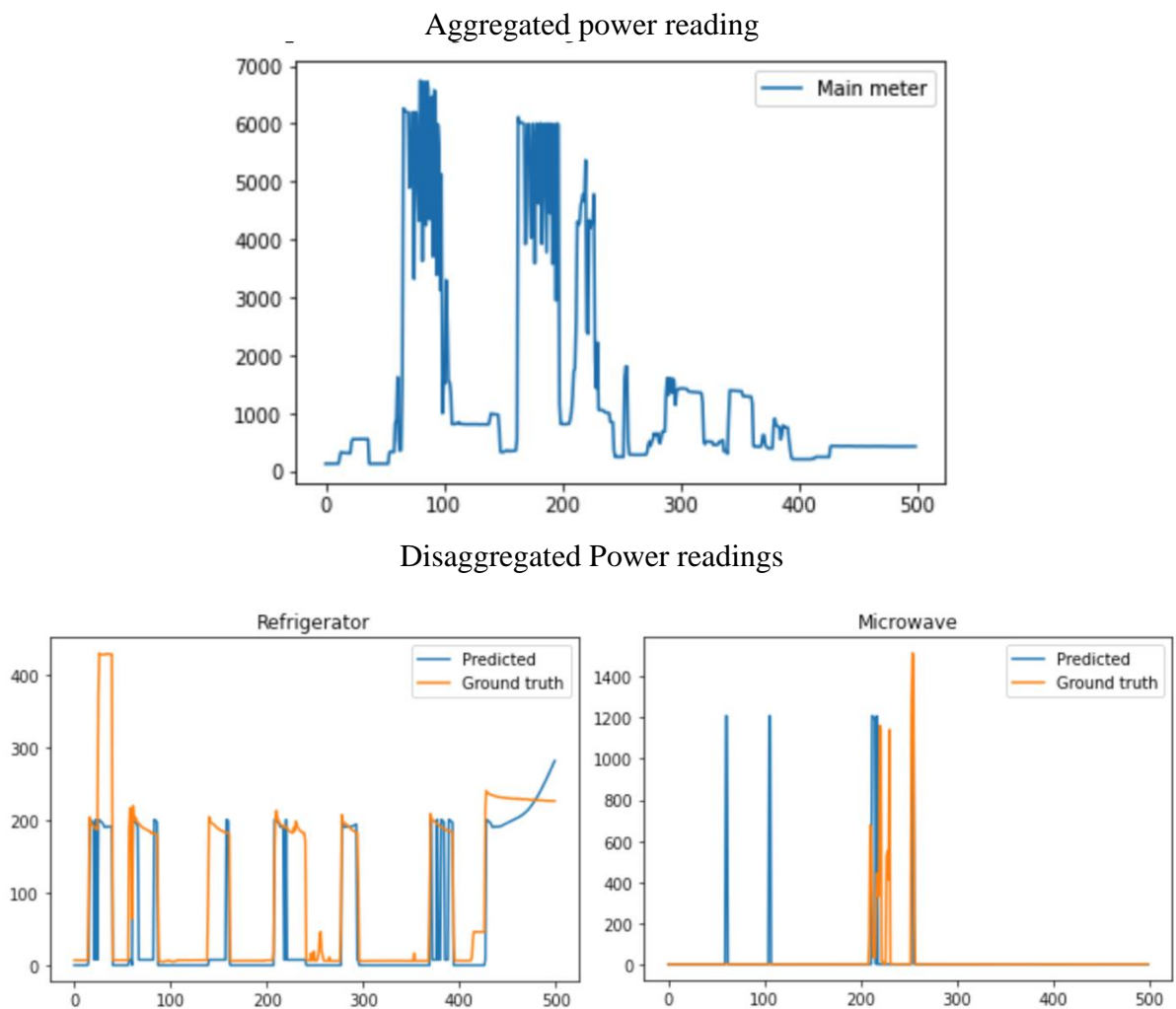

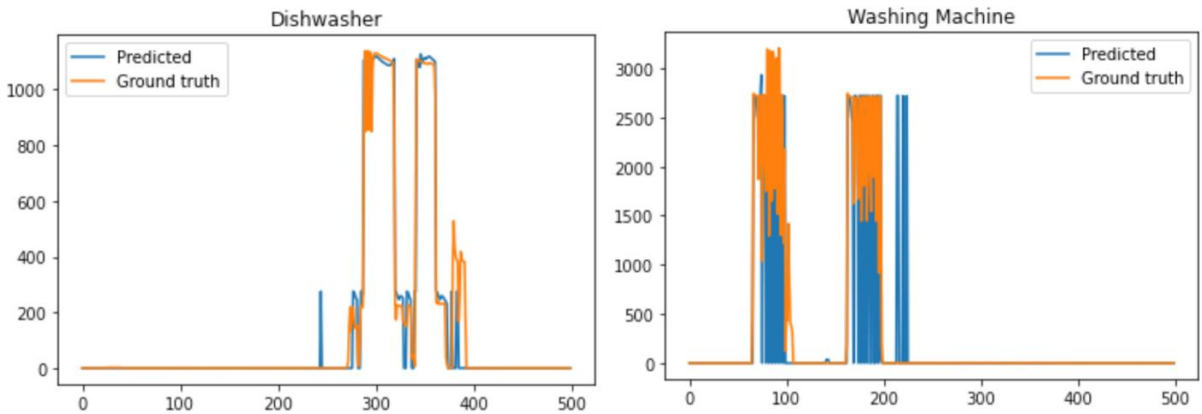

Figure 4 Disaggregation results of proposed method.

**CONCLUSION:**

In this paper, we have proposed a load disaggregation algorithm by using 1D-CNN and LSTM for low sampling data (1 min frequency) of smart meters. We have applied the proposed algorithm to a real-world dataset. We tested our approach and chosee three metrics to evaluate our model against the current state-of-the-art seq2seq, seq2point, GLU-res and CNN. The results show that our proposed method can correctly detect the operating state with the 97% accuracy and 0.91 F1 score on the same houses. On the unseen house the proposed method gives 12.3/9074 and 13.9/8144.5 MAE/MSE on power retrieval of microwave and dishwasher respectively. The results show that our proposed approach has good generalization ability, identification of operating state and power retrieval of multi-state appliance. Proposed method has 70K total parameters that consume 268 KB space to store the model weights; thus, our approach has the advantage of quick execution in the real-time application, requires lower memory space, has high accuracy compared to state-of-the-art and can be used on low cost embedded board such as ESP32 without any additional equipment for real-time feedback. The proposed method can give real-time appliance-specific energy feedback to the end-users which helps in identifying faulty devices and unwanted active appliances.